# Topology-enabled highly efficient beam combination

Yuhao Jing, [1,7] Yucong Yang, [2,3,7] Wei Yan, [2,3] Songgang Cai, [2,3] Jiejun Su, [2,3] Weihan Long, [4] Nuo Chen, [1] Yu Yu, [1,5,6] Lei Bi[2,3] & Yuntian Chen[1,5,6]

Beam combination with high efficiency is desirable to overcome the power limit of single electromagnetic sources, enabling long-distance optical communication and high-power laser. The efficiency of coherent beam combination is severely limited by the phase correlation between different input light beams. Here, we theoretically proposed and experimentally demonstrated a new mechanism for beam combining, the topology-enabled beam combination (TEBC), from multiple spatial channels with high efficiency based on a unidirectional topological edge state. We show that the topologically protected power orthogonal excitation arising from both the unidirectional edge states and the energy conservation ensures -0.31dB (93%) efficiency experimentally for a multi-channel combination of coherent microwaves at 9.1-9.3 GHz. Moreover, we demonstrate broadband, phase insensitive, and high-efficiency beam combination using the TEBC mechanism with one single topological photonic crystal device, which significantly reduces the device footprint and design complexity. Our scheme transcends the limits of the required phase correlations in the scenario of coherent beam combination and the number of combined channels in the scenario of incoherent beam combination.

[1]School of Optical and Electronic Information, Huazhong University of Science and Technology, Wuhan, 430074, China.
[2]National Engineering Research Centre of Electromagnetic Radiation Control Materials, University of Electronic Science and Technology of China, Chengdu 610054, China.
[3]State Key Laboratory of Electronic Thin-Films and Integrated Devices, University of Electronic Science and Technology of China, Chengdu 610054, China.
[4]School of Electronic Science and Engineering, University of Electronic Science and Technology of China, Chengdu 610054, China.
[5]Wuhan National Laboratory of Optoelectronics, Huazhong University of Science and Technology, Wuhan, 430074, China.
[6]Optics Valley Laboratory, Hubei 430074, China.
[7]These authors contributed equally: Yuhao Jing, Yucong Yang.
Correspondence and requests for materials should be addressed to Y. Y. (email: yuyu@mail.hust.edu.cn), L. B. (email: bilei@uestc.edu.cn), or to Y. T. C. (email: yuntian@hust.edu.cn)

**Introduction**

Power scaling of electromagnetic radiation is a demanding technology for many applications, including laser particle accelerators[1–6], advanced materials processing[7–10], and medical treatment[11]. In microwave frequency, the high-power systems[12] also have high potential applicability in areas such as directed energy, space-to-earth energy transfer[13], and high-power radar. Notably, the power scaling of the single laser systems has encountered several physical limitations[14–16], thus beam combination of multiple output power of laser systems has been the first measure to boost the power level. As for continuous waves, beam-combining technology can be categorized into coherent or incoherent beam combinations. The coherent beam combination essentially relies on the carefully delayed and locked phase of each combination channel for either tiled aperture or filled aperture geometries to maintain mutual coherence in both space and time.[17–24] Despite significant advances in recent years[25–33], this technology relies on complicated feedback and control systems. For monochromatic incoherent beam combinations, the passive combiner, such as polarization beam splitters[34], has a very limited number of input channels that maintain power orthogonality. In this regard, it is important to develop a beam combination strategy that can operate without complicated feedback and simultaneously admit multiple input channels.

In this work, we report an unprecedented strategy to realize a highly efficient electromagnetic beam combination enabled by topologically protected scattering-free edge states[35–40], namely the topology-enabled beam combination (TEBC). As sketched in Fig. 1(a), the principle of bulk-edge correspondence guarantees the existence of the unidirectional edge state, the direction of which depends on the relative values of topological invariants of the associated bulk materials. As such, one can pack the topologically different bulk materials in a spiral fashion by cascading down their Chern numbers, as shown in Fig. 1(b). Highly efficient beam combination can be realized for all the boundary states, since each input channel is scattering-free and only supports one forward propagating state. The edge states are robust against the material or structure imperfections as long as the band gaps are open. Thus, the electromagnetic power flows from N-input channels to a single output channel due to the absence of backward propagating states. The topologically protected power orthogonal excitation ensures that the output modes, excited by any input sources, are pairwisely orthogonal in power basis. We first illustrate our idea using a Y-shaped combiner formed by three topologically distinct photonic crystals, where the theoretical prediction perfectly agrees with numerical calculations. Based on the same underlying principle, we design and fabricate a 3×1 combiner containing four topologically distinct photonic crystals, with experimental beam combination efficiency upper to 93%.

**Results**

**Beam combination model based on spiral-staircase topology.** In topology-related physics, a significant result is the existence of gapless edge states localized at the interface as the topological invariant of two neighbor bulk materials changes. In two-dimensional (2D) systems, the topological invariant is usually coined as Chern number and calculated by integrating Berry curvature over the entire 2D Brillouin zone[41,42]. Implied by the stability of the topology associated with the Bloch modes across the entire Brillouin zone, the chiral edge state is robust against small perturbations, *i.e.*, without closing or opening the bandgap. In the topological scenario, the broken of time-reversal symmetry at the Dirac point brings a massive Dirac Hamiltonian, $\mathcal{H}(\mathbf{k}) = \hbar v_F \mathbf{k} \cdot \boldsymbol{\sigma} + m\sigma_z$, where the mass term $m$ flips sign across two topologically different bulk materials[38]. At the interface, $\mathcal{H}(\mathbf{k})$ has an elegant stationary solution given by $E(k_x) = \hbar v_F k_x$, where group velocity $v_F = 1/\hbar \cdot dE/dk_x$ and the sign of $v_F$ determines the chirality of edge states. Although $E(k_x)$ could develop a kink, the number of edge states is fully determined by the topological structure of the bulk states, which has been dictated by the bulk-boundary correspondence. Specifically, the gap Chern numbers of two neighboring bulk photonic crystals are $C_m$ and $C_n$ respectively, the difference and its sign, *i.e.*, $\Delta n = C_m - C_n$, essentially determines the number and propagation direction of edge states, as shown in Fig. 1(a).

The general principle of our topology-enabled beam combination is sketched in Fig. 1(b), which shows a spiral staircase distribution of gap Chern number of topologically different bulk materials. Each stair interface plays the role of one input channel and only supports one chiral edge state, while the cliff spatial channel acts as the output port, as indicated by the red and blue arrows, respectively. The total input energy is forced to combine at the output port, as guaranteed by both the chirality of edge states and the law of energy conservation. More importantly, the topological property for scattering-free edge states renders the beam combination phase-insensitive. The number of beam combination channels is determined by the difference in the gap Chern number across the cliff interface. It shows the advantage against the incoherent beam combination between orthogonal polarized electromagnetic waves, which is usually limited to two channels.

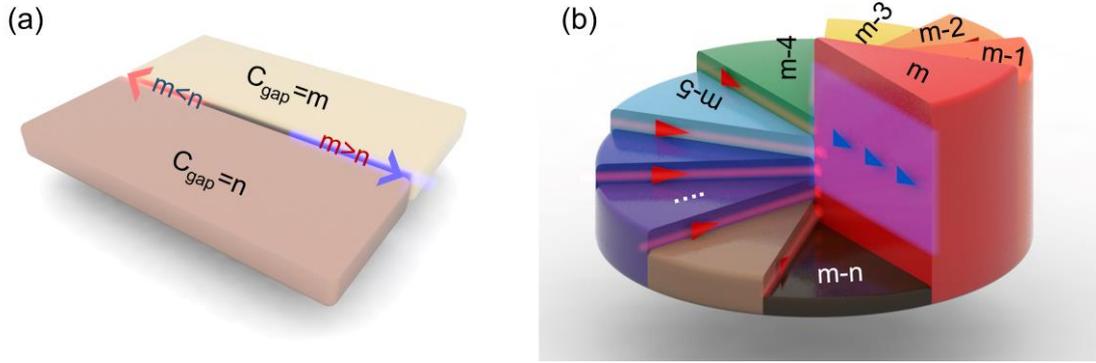

**Fig. 1 The chiral edge states and the spiral staircase distribution of Chern number. a** The number and propagation direction of the edge states are determined by the absolute value and the sign of $C_m - C_n$ of two neighboring bulk materials, respectively. The band gaps of the bulk materials are overlapped in frequency. The edge states propagate towards left/right, as indicated by the red/blue arrow, if the gap Chern number of the yellow region is smaller/larger than that of the red region. **b** The spiral-staircase topology of gap Chern numbers with the fan-shaped distribution. Each sector represents a non-trivial topological material with a gap Chern number indicated by the black/white number. At all stair interfaces, the difference in gap Chern number between the two neighboring regions is one and the edge states propagate inwards. In contrast, the difference of gap Chern numbers is n and the edge states propagate outward at the cliff interface.

**Experimental implementation of the spiral-staircase topology.** Inspired by recent progress in topological photonics, the spiral-staircase topology for beam combination can be realized simply by assembling photonic crystal based on Chern insulators properly[37,40]. Generally, the large Chern number requires the participation of more pairs of Dirac points or quadratic points gapped by time-reversal symmetry breaking. Both of them could be achieved by symmetry-protected degeneracy or accidental degeneracy. The former is related to geometry symmetry, while the latter needs precise parameters design. Therefore, to illustrate the basic idea of the spiral-staircase topological device, we assemble four types of photonic crystal with different gap Chern numbers together, as shown in Fig. 2(a). As sketched in Fig. 2(b), the four various photonic crystals share a similar unit cell structure, which contains the yttrium iron garnet (YIG) rod sandwiched between two copper-clad laminates (CCLs, see Method for details). Each YIG rod is biased by a NdFeB magnet-provided magnetic field. The magnetized YIG-based photonic crystal generates the complete band gaps for TM polarization at microwave frequencies. Specially, we utilized four mm-height YIG rods with the radius of $r_1$, $r_2$, $r_3$, and $r_4$ illustrated in Fig. 2(c-f) to construct photonic crystals with different gap Chern numbers ($C_{gap}=2$, $C_{gap}=1$, $C_{gap}=0$, $C_{gap}=-1$) in a frequency range of 8.95-9.45 GHz with overlapped bandgap.

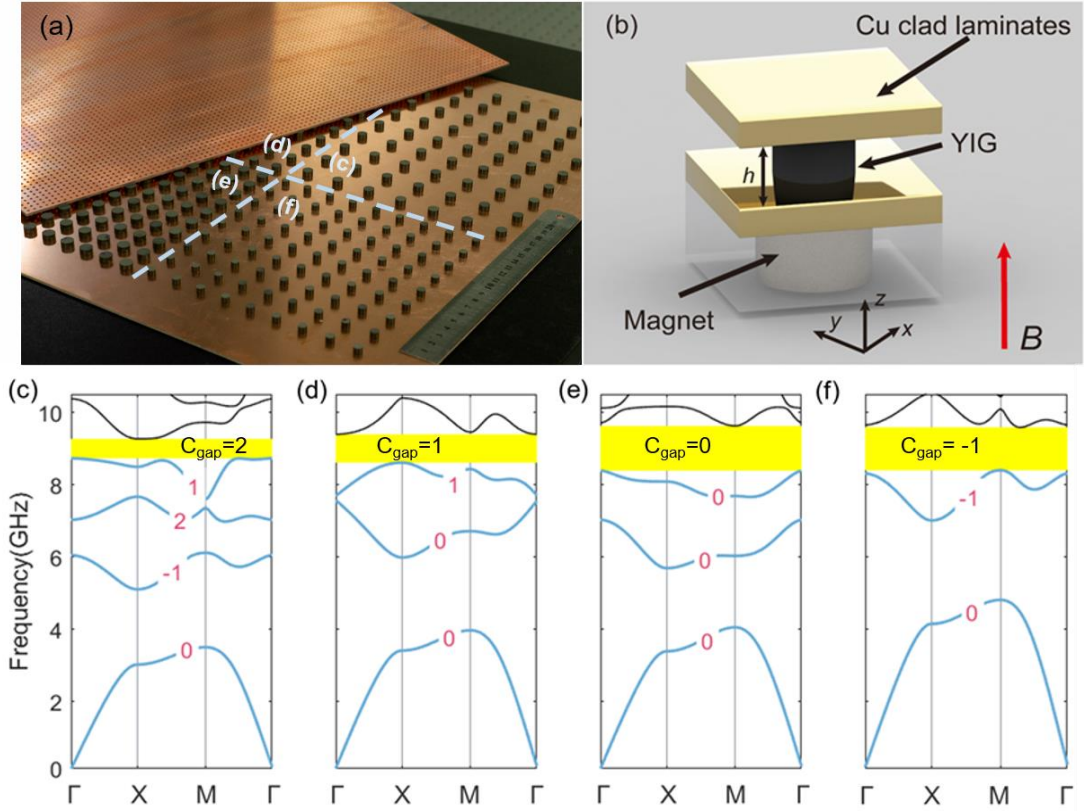

**Fig.2 Topological non-trivial photonic crystal structures and photonic band structures. a** Photonic crystal with YIG rods arranged in a square lattice with variable radius and lattice constant. **b** The unit cell of the photonic crystal. **c ~ f** The complete photonic band structure for TM polarization. The Chern number of each band is indicated by the red number. The gap Chern number can be calculated by summing the Chern numbers of each band below the band gap. Regions of band gaps are highlighted in yellow.

We proceed to discuss the experimental verification of the topology-enabled beam combination by measuring the real-space field distribution[43], as shown in Fig. 3(a). To probe the mode profile, we placed three coherent dipole antennas at three marked positions, labeled by orange stars in Fig. 3(b), and moved the probe antennas to record the magnitude and phase of the electric field at each position (see Method for details). Fig. 3(b) shows the measured distribution of $E_z$ component at 9 GHz. We also measured the field profile for the input frequency range of 8-10 GHz. The edge states unidirectionally propagate along the interfaces between photonic crystals with different Chern numbers, preserving fairly good field confinement.

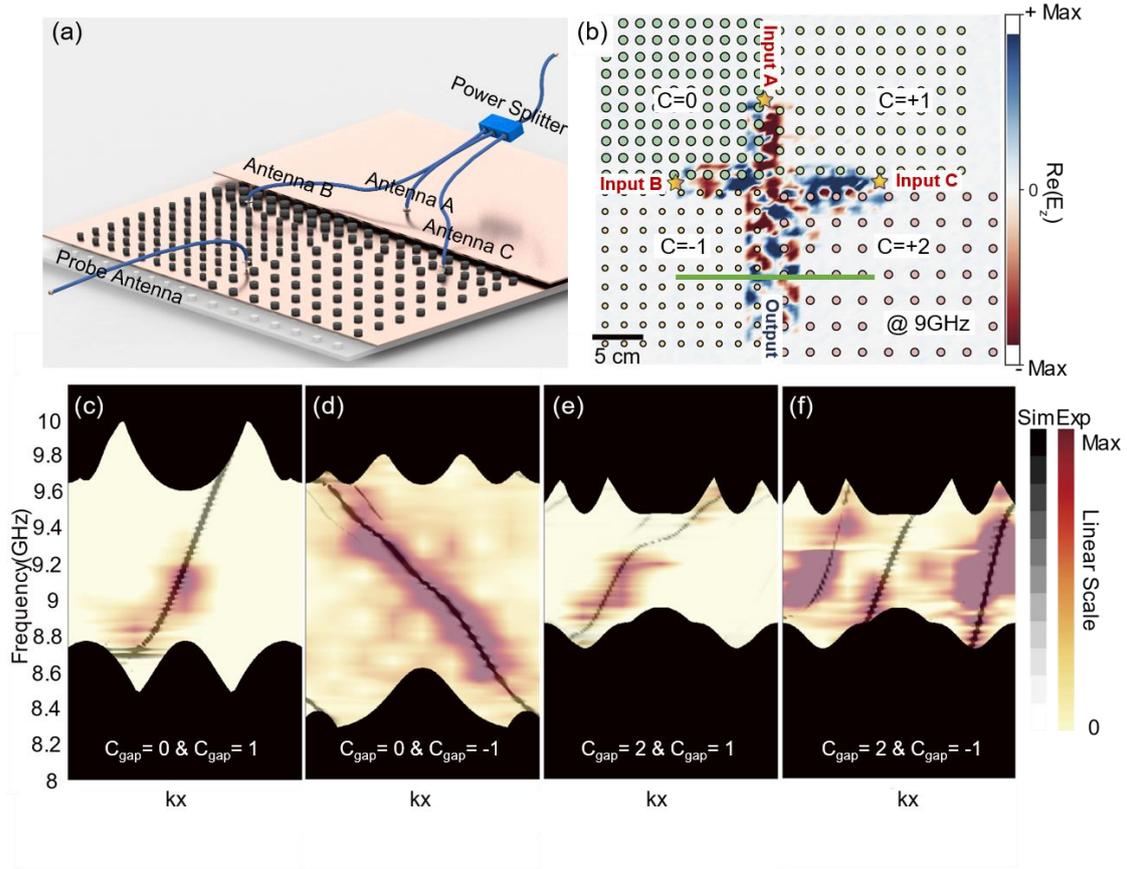

**Fig. 3 Experimental setup, measured field profile for beam combination and the projected band. a** Schematic of the fabricated YIG photonic crystal sample. **b** Measured $Re(E_z)$ distribution in the photonic crystal for three input sources at 9 GHz frequency, demonstrating topology-enabled beam combination. The green line represents the location of the line integral used in power orthogonal excitation. **c ~ f** The projected band for four spatial channels. The FT of mode profile directly demonstrate the dispersion of top ($C_{gap} = 0$ and $C_{gap} = 1$, Input A), left ($C_{gap} = 0$ and $C_{gap} = -1$, Input B), right ($C_{gap} = 2$ and $C_{gap} = 1$, Input C) and bottom ($C_{gap} = 2$ and $C_{gap} = -1$, Output) spatial channels respectively. The FT from simulation and experiment data are emphasized by the dark-white color bar and red-yellow color bar separately.

Figure 3(c-f) shows the projected band structure obtained by converting the measured and simulated mode profile of the edge states from real space to k-space via discrete Fourier transform (FT) in a frequency range of 8-10 GHz. The dispersion of edge states obtained from the discrete FT explicitly shows separate spatial channels located inside the overlapped bandgap in Fig. 3(c-f). Moreover, the peak location of the measured dispersion of the edge states matches perfectly well theoretically-calculated results. The sign of group velocity for those edge states could be obtained from Fig. 3 (c-f), indicating the unidirectional energy flow predicted in the spiral-staircase topology.

Compared to the calculated band structure, the linewidth of the edge states from the experiment is wider, which is due to the measurement accuracy limited by the finite sampling and the small lattice constant (5mm) of the holes array of the CCLs. The uncompleted band gap and discontinuity of the experimentally measured edge state are originated from fabrication-caused inhomogeneity of the applied magnetic field.

**Phase-insensitive high efficiency beam combination.** In conventional beam combination optical systems, complicated feedback and optical structures have been adopted to build the phase correlation precisely between different input beams[31,44]. In contrast, the TEBC has no such need for controlling phases among different channels, thus showing certain advantages in terms of simplicity and efficiency. Next, we discuss the efficiency of topology-enabled beam combination. To that end, we fix the position of the probe antenna for the output channel and measure beam combination efficiency against the variation of (1) the operation frequency varies and (2) the position of excitation antennas (see Method for details).

We first measured the beam combination efficiency as a function of the operation frequency. The yellow region in Fig. 4(a) represents the overlapped bandgap region of the four photonic crystals. Inside the yellow region, all the edge states exhibit strong unidirectionality. As for the forward propagating states (labeled in red line), the beam combination shows high transmission in the band gap frequency range. Whereas the backward propagation (labeled in blue line) loss is up to 20~55 dB, indicating that the backward propagating state is forbidden. At the frequency of 9.12 GHz, the forward transmission loss at the output port is only -0.31 dB, which is primarily due to the material absorption of YIG. In the frequency range of 9 to 9.1 GHz which is close to the bulk band, the transmission is slightly dropped due to the fact that the excitation efficiency is not exactly 100%. In the frequency region between 9.25 GHz and 9.45 GHz, beam combination efficiency drops significantly due to the inhomogeneity of the applied magnetic field, leading to a narrower bandgap than that used in the simulation. See details about the impact of inhomogeneity of the applied magnetic field in the supplementary.

We further measured beam combination efficiency as a function of the operation frequency and the position of the excitation antenna as plotted in Fig.4 (b), wherein the coordinate of the excitation antenna varies from 0 cm to 7 cm (0-λ, λ is the vacuum wavelength at 9 GHz). In the frequency range of 8.98-9.25 GHz, the transmission loss is lower than -1 dB. The variation of transmission loss is below 0.8 dB as the dipole antenna moved across a whole process at 9.1 GHz. Remarkably, the variation of the position of the excitation antenna will change the phase correlation among three input edge states at the cross junction, wherein the high efficiency of beam combination outright demonstrates topologically protected phase-insensitive.

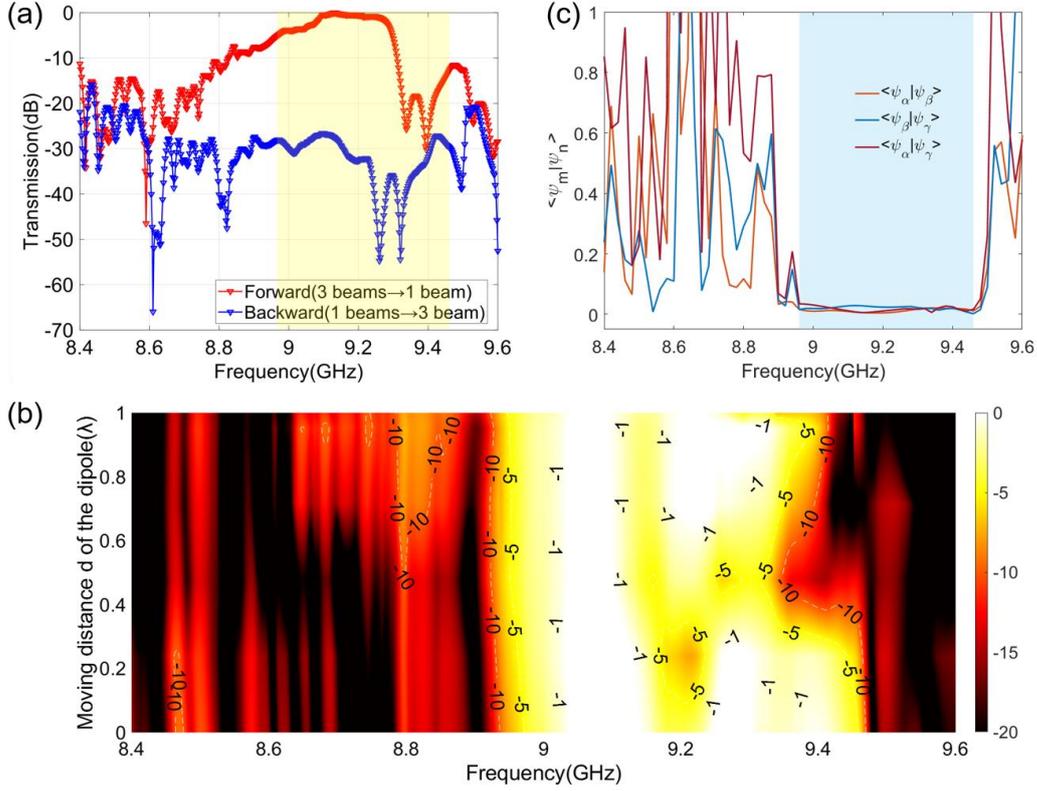

**Fig. 4 Phase-insensitive highly-efficient topology-enabled beam combination. a** The experimental transmission spectra for forward (3 beams to 1beam) and backward (1 beam to 3 beams) propagation. **b** Experimental transmission of forward (3 beams to 1 beam) as a function of the frequency and the moving distance of the dipole. **c** Orthogonality of topological boundary modes. The three input modes are marked as $\psi_A, \psi_B, \psi_C$, when only the dipole located at input A, B or C spatial channel is active. All three groups of inner products are close to zero in their shared band gap highlighted in blue, which shows good orthogonality among the three excited modes.

**Topologically protected power orthogonal excitation.** The spatially overlapped high dimensional modal space is only a necessary condition to realize highly efficient beam combination, while topologically protected unidirectionality of the propagating edge states spanning a high dimensional modal space in junction with energy conservation is a sufficient condition. Indeed, our proposal of efficient beam combination is due to energy conservation in junction with the topology-protected unidirectionality of those edge states, and can also be interpreted from classical electromagnetics in terms of "power orthogonal excitation". Namely, for a given source launched at any input channel

$i$, the excited state ($\psi_i$) at the cliff interface, i.e., a superposition of multiple edge states, shall be orthogonal to that ($\psi_j$) of input channel $j$. Mathematically, the power orthogonal excitation can be described[45] as follows,

$$\langle \psi_i | \psi_j \rangle = \frac{\int \left(E_j \times H_i^* + E_i^* \times H_j\right)_y dl}{\sqrt{\int (E_i \times H_i^* + E_i^* \times H_i)_y dl \cdot \int (E_i \times H_i^* + E_i^* \times H_i)_y dl}} = \delta_{ij}$$

where $E_i/H_i$ ($E_j/H_j$) is the electric/magnetic field component associated with the excited state $\psi_i$ ($\psi_j$), $\delta_{ij}$ is the Kronecker symbol, and the line integral could locate at arbitrary cross section shown in Fig. 3(b). As for the 3×1 beam combiner in this work, the three excited states denotes as $\psi_i$, $i \in (\alpha, \beta, \gamma)$, where $\alpha$, $\beta$ and $\gamma$ represents individual excitation from the three different input channel. Three excited states are pair-wisely orthogonal, as evident from the projection $\langle \psi_i | \psi_j \rangle$ in the frequency range of 8.95 to 9.45 GHz, as shown in Fig. 4(c). The power orthogonal excitation indicates that the three excited states, i.e., $\psi_i$, $i \in (\alpha, \beta, \gamma)$, are mutually independent and thus fulfill the superposition principle in terms of optical power full rather than the optical field, which bears a resemblance to the incoherent beam combination between two orthogonal polarized light beams. In contrast, topology-enabled beam combination has no limit to the number of combining channels, as long as the requirement of power orthogonal excitation is fulfilled.

A rigorous description of the power orthogonal excitation can be given using a scattering matrix; see details in supplementary. The idea is to take the absence of the scattering channels to accomplish the dimensional reduction of the scattering matrix at the cross junction where the beam combination occurs. Thus, the three edge states located at three separate spatial channels could be regarded as the input ports for the scattering matrix, while the output ports are three edge states lying on the same spatial channel. Under the requirements of unidirectional transport and energy conversation, as explained in Supplementary Note 5, the power orthogonal excitation can be seen as a consequence of the hermicity of the scattering matrix.

**Discussion**

We propose the concept of topology-enabled beam combination (TEBC) and design a compound structure with spiral-staircase topology, to realize efficient beam combination. The proof-of-concept of TEBC has been verified in both numerical simulation and experimental measurement, showing beam combination efficiency upper to 93% over a broad frequency range of 9.1-9.3 GHz, which is useful in long-distance wireless energy transmission and high-power radar. We have also experimentally proved that our proposed TEBC maintains high efficiency and broad bandwidth against phase variations of any combination channel. From the viewpoint of classical electromagnetics, we proposed the concept of power orthogonal excitation to place our TEBC in the context of incoherent beam combination based on polarization and further expand it in a more generic scenario, wherein multiple spatial overlapped yet power orthogonal channels in incoherent beam combination can be handled.

In addition, the operating frequency of TEBC can also be extended to optical and telecommunication frequencies. For on-chip integrated, efficient beam combination, we can use yttrium iron garnet thin films with permittivity tensors such as cerium or bismuth-doped yttrium iron garnet, which can be monolithically deposited on silicon or silicon nitride substrates to form TE-mode photonic crystals[46,47]. In this scenario, we can achieve spiral-staircase topology of multiple photonic crystals by applying a magnetic field to break the time-reversal symmetry and construct an on-chip integrated topological beam-combining device. On-chip integrated, efficient beam combination can be demonstrated by careful design and continuous material improvement.

## Method
### Numerical simulation.

The full-wave results of the band structures and mode profiles shown in Figs. 2, 3 and 4 are performed using the commercial finite element method software COMSOL Multiphysics, in which Electromagnetic waves, Frequency domain (EWFD) module is used to demonstrate topological non-trivial TM band structures. In order to calculate the Chern number of each band below the interest band gap, a gauge-invariant numerical method has been adopted to calculate the band Chern number in discretized Brillouin zone.

When calculating the excited edge states mode profile, the magnetic dipoles are placed to excite edge states located at different spatial channels. Perfectly matched layer is used to properly truncate the simulation domain. One dimensional discrete Fourier Transform is adopted to produce the dispersion of edge states.

### Device fabrication

Topology-enabled incoherent beam combination device is composed of square arrays of YIG pillars with different radius and lattice constants, sandwiched in between two 500 mm× 500 mm copper clad laminates (CLLs). The radius of the four photonic crystals are 4.172 mm, 4.054 mm, 4.748 mm, 3.038 mm respectively, and the corresponding lattice constants are 2.7814 cm, 2.1452 cm, 1.8054 cm, 2.0253 cm, and all YIG pillars are set to 4 mm in height to match the distance between CLLs (as shown in Fig. 2a). The two CCLs are placed to face-to-face and used as the reflectors to form a Fabre-Polar cavity. In the plane, photonic crystal arrays are placed in the same pattern as sketched in Fig.2a, leading to four interfaces separating four topologically different bulk materials. YIG pillars are made of YIG block crystal using ultra-high precision computer numerical control (CNC) machine (Mazak variaxis i700), with a processing error of less than 0.05 mm.

In order to firmly fix the YIG pillars on the copper-clad laminated substrate, double-sided tape is first applied to the bottom side of YIG pillars. Subsequently, each YIG pillar is aligned with precisely defined acrylic mold with holes array, the radius, lattice constant and height have one-to-one corresponding to that of YIG pillars). After pressing the YIG pillars with double-sided, and accurately fixing the YIG pillars onto the CCL bottom plate through the mold, we carefully removed the acrylic mold to prevent damage to the YIG pillars.

Firstly, a 2mm thick acrylic plate with double-sided tape is set to form the substrate, which is used to align the NdFeB magnets accurately underneath each YIG column, and to make sure appropriate magnetization of each YIG pillar. Then, a 5mm thick acrylic mold with precisely defined configuration is fixed on the prepared substrate. At last, the NdFeB magnets are settled in the holes array of the 5mm thick acrylic plate with the same direction, which provides the applied magnetic field.

**Characterization Setup**

We use four antennas setup, *i.e.*, three dipole antennas as the transmitter and one dipole antenna as the receiver, to characterize our samples. The input electromagnetic wave is connected to a MNK Communication power splitter (Working Freq:5 GHz-10GHz) through a coaxial cable. The input electromagnetic wave is divided into three coherent electromagnetic waves of the same frequency, while the output electromagnetic wave is received by the probe dipole antenna. The projected band structure, transmission spectra and near-field distribution of our proposed TEBC device are measured using our four antennas setup. The four antennas are inserted into the waveguide through holes drilled with an ultra-high precision CNC machine and connected to the vector network analyzer (rod and Schwartz ZNB 20) to measure S-parameters. In all the measurements, the 3.5 mm 85052D through-off-short-match (TOSM) calibration and short-circuit calibration of the dipole antenna are performed, hence the measured S-parameters only contain the insertion loss of the device.

**Acknowledgement**

The authors are grateful for support by National Natural Science Foundation of China (NSFC) (Grant No. 12274161, 11874026, 51972044 and 52021001), the Ministry of Science and Technology of the People's Republic of China (MOST) (Grant No. 2018YFE0109200 and 2021YFB2801600), Sichuan Provincial Science and Technology Department (Grant No. 2019YFH0154 and 2021YFSY0016), Fundamental Research Funds for the Central Universities (Grant No. ZYGX2020J005) and the Innovation Project of Optics Valley Laboratory.